\documentclass[manuscript,screen,nonacm]{acmart}
\AtBeginDocument{%
  \providecommand\BibTeX{{%
    \normalfont B\kern-0.5em{\scshape i\kern-0.25em b}\kern-0.8em\TeX}}}

\usepackage{hyperref}
\usepackage[
    type={CC},
    modifier={by-nc-sa},
    version={4.0},
]{doclicense}

\author{Camilo Sanchez}
\orcid{0000-0002-8486-031X}
\title{}
\email{camilo.sanchez@aalto.fi}
\affiliation{%
  \institution{Aalto University}
  \city{Espoo}
  \country{Finland}
}

\author{Felix A. Epp}
\orcid{0000-0001-6252-7244}
\title{}
\email{felix.epp@aalto.fi}
\affiliation{
  \institution{Aalto University}
  \city{Espoo}
  \country{Finland}
}

\begin{document}
\doclicenseThis
\title{Experiential Futures In-the-wild to Inform Policy Design}






\maketitle

\section{Messy Experiential Futures}
Technological innovation shapes the world around us at an accelerating pace. 
By building novel technologies, HCI materialises visions of the future in the present \cite{alvial-palavicinoFuturePracticeFramework2016, gustonDaddyCanHave2013, salovaaraScalingHCIPrototype2020}. 
The unintended consequences of these technologies then require policy makers to attend to them retroactively.
However, policy development has been struggling to keep up with the pace of the innovation cycles, which increases uncertainty and limits evidence-based regulation \cite{owen2013framework}.
To address the policy-novelty gap, R. von Schomberg proposed the Responsible Research Innovation (RRI) to drive science and technology innovation towards socially desirable goals and respond to the "grand challenges" of our time \cite{vonschombergVisionResponsibleResearch2013}.
While adopting the RRI poses challenges in terms of scale, temporality, and resources \cite{grimpeCloserDialoguePolicy2014a}, HCI can offer participatory, sustainable, and value-laden perspectives \cite{batesResponsibleInnovationAgenda2019} that policy designers could adopt to operate on the dimensions proposed by RRI \cite{owen2013framework}.
To better understand the impacts of technology in long-term and uncertain futures, HCI should consider other driving forces of society beyond the technological~\cite{salovaaraScalingHCIPrototype2020, mankoffLookingYesterdayTomorrow2013}

Future studies established that considering futures as uncertain and open-ended can help identify fluctuating knowledge still subject to change~\cite{adamFutureMattersAction2007, andersonPreemptionPrecautionPreparedness2010}, which relates to RRI by anticipating emerging socio-ethical concerns to alter policy goals \cite{sykesResponsibleInnovationOpening2013}.
Anticipation can be exercised through scenario planning, prototypes, life cycle assessments, role-playing and other methods~\cite{alvial-palavicinoFuturePracticeFramework2016} to adjust our assumptions and relations to possible futures~\cite{andersonPreemptionPrecautionPreparedness2010}.
As future narratives may reflect biases 
not representative of diverse socio-cultural groups \cite{kinsleyFuturesMakingPractices2012, lightCollaborativeSpeculationAnticipation2021, mazePoliticsDesigningVisions2019}, futures studies scholars are developing more participatory methods 
\cite{candyTurningForesightOut2019, millerTransformingFutureAnticipation2018, dahle50KeyWorks1996}.
An emerging collection of participatory methods is Experiential Futures, which addresses the predominance of expert-based textual construction of futures by engaging socially diverse stakeholders in the future-making process through tangible, performative and interactive experiences of everyday life visions~\cite{gardunogarciaDesigningFutureExperiences2021, candyTurningForesightOut2019, jenkinsFutureSupermarketCase2020}. 


Akin to adopting HCI methods in Responsible Innovation~\cite{batesResponsibleInnovationAgenda2019}, HCI provides opportunities to incorporate everyday life experiences and participatory approaches in foresight. 
Considering ubiquitous computing, HCI developed capacities to study the "messiness"~\cite{bellYesterdayTomorrowsNotes2007} of everyday life, which provides useful avenues for experiential futures.
Field trials and design interventions study socio-political and cultural factors in users' ordinary life to examine the understanding and adoption of technologies~\cite{brownWildChallengesOpportunities2011, eppAdornedMemesExploring2022, salovaaraScalingHCIPrototype2020}. 
These field interventions as experiential futures may provide a new opportunity for policy design.

\section{Case Study: Data Leakage in Sustainable Smart Garment Futures}
To this end, we are developing and conducting a field trial study, which fosters the idea of anticipatory practice~\cite{alvial-palavicinoFuturePracticeFramework2016}. 
Using the Future Ripples method \cite{eppReinventingWheelFuture2022}, we explored the trajectories of wearable technology futures from diverse STEEPLE factors (societal, technological, economical, ethical, political, legal, and environmental). Consequently, we identified the collision between data privacy and recycled electronics in the context of the Circular Economy.

While our study aims to materialise an experiential future in-the-wild, we iteratively developed future scenarios in parallel to the concretisation of an interactive prototype. 
Similarly, to the “multiply” step in Ethnographic Experiential Futures \cite{candyTurningForesightOut2019}, the initial scenario was refined into four future scenarios by using consequences mapped out in five Future Ripples workshops and identifying the critical uncertainties~\cite{vanderheijdenScenariosArtStrategic2005} of surveillance and social divide.
The four scenarios, enriched with weak signals and trends, 
guided us to reiterate the prototype and field intervention. 

By placing ubiquitous technology in the form of a tracking garment and a stationary device for reflections in the home for a 2-week period, we confront participants with a future scenario of data leakage through recycled electronics. This way, we aim for a pervasive experiential future that taps into the routines, values, and meanings that people give to their everyday interaction with technology before embedding it into society~\cite{wongElicitingValuesReflections2017, grimpeCloserDialoguePolicy2014a}. 
The insights we hope to learn from the field intervention and reflections on the futures should inform policies regarding possible unintended consequences and peoples' values, fears, and assumptions. 

\section{Proposition for Further Inquiry}
Although there have been proposals in HCI to attend socio-political reflections on future visions through experiential futures \cite{jenkinsFutureSupermarketCase2020} and counterfactual actions \cite{forlanoSpeculativeHistoriesJust2023}, there is still much ground to empirically research HCI's role in the materialisation of collective future visions. 
This work could be a dualistic endeavour between policy design and HCI. On the one hand, policy design could benefit from more agile and responsive evaluation environments. On the other, HCI studies can improve the impact of their findings beyond their own discipline. 
For both fields, engaging in pluralistic futures rendering means questioning which and how futures are constructed, who are they benefiting, and how are the findings of these interventions interpreted towards other futures? 
For HCI's cross-disciplinary quality, adopting futuring methods to aid policy design should not be an upstream task.  
However, HCI itself should give more weight to its own capacity to study futures empirically and the responsibility this entails. 

\begin{acks}
This work has been supported by the Academy of Finland grant Future Methods (330124).
\end{acks}

\bibliographystyle{ACM-Reference-Format}
\bibliography{manuscript}

\end{document}